\documentclass[runningheads]{llncs}
\usepackage[utf8]{inputenc}
\usepackage{todonotes}
\usepackage{listings}

\title{DApp for Rating}

\author{Andreea Buțerchi\inst{1}, Andrei Arusoaie\inst{1}}
\institute{Alexandru Ioan Cuza, University of Ia{\c s}i, \email{andreea.buterchi,arusoaie.andrei@info.uaic.ro}}
 \date{July 2020}
\authorrunning{Bu{\c t}erchi et. al.}

\begin{document}
\maketitle

\begin{abstract}
Lots of existing web applications include a component for rating internet resources (e.g., social media platforms include mechanisms for rating videos or posts). 
Based on the obtained rating, the most popular internet resources can generate large amounts of money from advertising. One issue here is that the existing rating systems resources are entirely controlled by a single entity (e.g., social media platforms). 

In this paper we present a blockchain-based decentralized application for rating internet resources. The proposed solution provides a transparent rating mechanism, since no central authority is involved and the rating operations are handled by a specialised \emph{smart contract}.
We provide an implementation of our idea, where we combine existing authentication methods with blockchain specific features so that anonymity is preserved. We show that this approach is better than existing rating components present in various web applications.
\end{abstract}

\section{Introduction}

These days, blockchain-based decentralized applications are designed for a variety of use cases. The society embraced the innovation of the blockchain technology and transformed its potential into ingenious software products. Since one of the main purposes of the blockchain technology is to remove the interaction with a central authority or a third party which has control over specific information, the potential of the blockchain was widely acknowledged and it was adopted in various domains: finance, government, digital identity, media and entertainment, healthcare and supply chain management. 

Rating systems are widely used by many social media platforms to establish specific trends or trending posts, videos, movies, and so on. There are various rating systems, which have different purposes with respect to the rated resources: \emph{like/dislike-based rating systems}, \emph{star-based rating systems} and \emph{review-based rating systems}, which are used for media contents and products evaluation.

Most of the existing rating systems are used for advertisement, which can provide the one who controls a specific resource with substantial incomes. An interesting fact is that most of the media platforms control directly the rating systems they are using. Even if the politics of media platforms seem to be transparent, there is still no evidence for this transparency. The users on these platforms have to trust the rating process, without any certain proof that it is indeed transparent.

To our best knowledge, at this moment there is no solution that has the purpose of moving rating systems into the blockchain. However, we can associate \emph{rating} with \emph{voting}, which is a concept widely discussed~\cite{10.1007/978-3-642-02627-0_5,mci/Khader2012,Ayed2017ACS,10.1007/978-981-10-7605-3_50,8603050}. In this case, the blockchain technology is used to distribute an open voting record among citizens, such that the citizens do not need anymore to put their trust into central authorities.

Considering the similarity between \emph{decentralized voting}  and \emph{decentralized rating}, we can admit that a blockchain-based decentralized application for rating could have a major impact, since it allows the users to rate internet resources in a trustworthy manner. 

Most of the existing media platforms include a component for rating their content (e.g., posts, videos, songs, movies). For example both YouTube and IMDb provide rating mechanisms. The rating process on these two platforms is handled by specialised implementations. 
Even if YouTube informs its users with respect to the number of views, likes or dislikes that a video resource collected, there are other factors that influence these numbers (e.g., watch time, session time or the popularity of a specific channel). Since the number of views collected by the video resources is usually a criteria for producing advertising money, the owners of these resources may be determined to use fraudulent means to increase their incomes. On the other hand, YouTube claims that their algorithms handle these kinds of situations, even if there is still a lack of transparency in their actions\footnote{There are plenty of web pages where users complain (e.g.,  \url{https://www.authorsguilds.com/how-youtube-algorithm-works/},
\url{https://support.google.com/youtube/thread/11131851?hl=en}, 
\url{https://www.appypie.com/how-youtube-algorithm-works}). Even if these pages cannot be necessarily trusted, it is hard for YouTube to guarantee/prove that its rating algorithms are transparent and work correctly.}.

On the IMDb platform, one of the main problems is the determination of the ranking of a specific movie. The platform does not consider the mean of the ratings, but it proposes a complex algorithm which computes the ranking based on some criteria. Even if the rating values are accessible, the users can not trust entirely the accuracy of the rating process, since its transparency is still under question \footnote{ \url{https://help.imdb.com/article/imdb/track-movies-tv/ratings-faq/G67Y87TFYYP6TWAV\#}.}.

In order to address these issues, we propose a decentralized application for rating. 
Besides the rating functionalities, which are provided by a specialised \emph{smart contract}, our solution comes with an additional authentication mechanism, that guarantees the \emph{uniqueness} of each user who rates resources. In this way, the fraud tendencies are reduced and the rating process is transparent and trustworthy.

\paragraph{Contributions.} The main contribution of this paper is a blockchain-based decentralized application for rating different kinds of internet resources. The application provides its users with an intuitive UI, which facilitates the rating process. The rating functionalities are implemented by a specialised \emph{smart contract}, which guarantees the protection of the identities of the users and stores rating records in a proper manner. 
    
\paragraph{Paper Organisation.}
In Section~\ref{sec:prelim} we present background information on the blockchain and the technologies we use. In Section~\ref{sec:tool} we present our main contribution, that is, a decentralized application for rating.
In Section~\ref{sec:tooleval} we present some experiments that we have performed and their execution cost analysis. We conclude in Section~\ref{sec:conclusions}.

\section{Background and tools}
\label{sec:prelim}

This section recalls several background notions and tools that we use in this paper. We give a brief presentation of each of them, and we point the readers to additional material if necessary.

The \emph{blockchain} is a new technology, which records data across a peer-to-peer network in a distributed shared ledger. The nodes of the network contribute to the creation this distributed ledger, which is chain of blocks of transactions. 
The creation of blocks is a complex process that requires computing power. Specialised nodes, called \emph{miners}, are incentivised to invest computational power in order to create new blocks.
Transactions are broadcasted in the network, and miners collect them and try to build a block of transactions following strict rules, including a link to the last known block in the chain of blocks. When a miner finishes the creation of a block, the block is send across the network and consensus algorithms are used to accept the block. The centralised control is out of the question. 
 Based on the consensus principles, approved blocks are added to the main ledger and stored chronologically. Even if this technology was initially designed for cryptocurrency trading~\cite{bitcoin}, its potential is used in a variety of software products.

\emph{Ethereum}~\cite{wood2014ethereum,eth} is a popular blockchain platform, which enables the use of \emph{smart contracts}. A smart contract is just a program that encodes a digital agreement which is executed automatically by the \emph{miners} in the network. Such programs are deployed in the network as special transactions and the miners need to execute them when creating blocks.
Smart contracts can simulate real-world agreements for different kinds of assets. Usually they are written in higher-level languages (e.g., Solidity~\cite{solidity}, Vyper~\cite{vyper}) and compiled into Ethereum bytecode~\cite{wood2014ethereum,eth}. The compiled code is packed in a transaction and sent across the network. Miners use the \emph{Ethereum Virtual Machine} (EVM) to execute compiled contracts.
Each bytecode instruction has an associated execution cost. This is yet another incentive for the miners to execute the code. On the other hand, the smart contract programmer needs to balance the benefits with the execution costs.

\emph{Solidity}~\cite{solidity} is a statically-typed, high-level and contract-oriented programming language designed for implementing smart contracts that run on the EVM. Solidity has a wide range of functionalities inspired by existing programming languages (e.g., JavaScript, C++ and Python). This programming language is used on several platforms, including Ethereum and Hyperledger~\cite{hyperledger}.
    
\emph{Remix}~\cite{remix} is a powerful tool that allows its users to design, deploy and test smart contracts directly in the browser. It provides development environments for several contract-oriented programming languages (e.g., Solidity, Vyper). Since Remix can simulate the deployment process of a smart contract, users can interact with the smart contract functions, emit events, debug or even inspect transactions. Users can choose from a wide range of plugins, which are designed to improve the development process of smart contracts (e.g., vulnerabilities detection, costs estimation, oracles integration). Remix can connect with Metamask~\cite{metamask} and deploy the smart contracts on various existing test networks (e.g., Ropsten~\cite{ropsten}, Kovan~\cite{kovan}, Rinkeby~\cite{rinkeby}).

\emph{Metamask}~\cite{metamask} is a browser extension, which can be used on various existing browsers (e.g., Chrome, Firefox and Brave). It represents a way to connect normal browsers with the Ethereum blockchain and also to interact with decentralized applications. Metamask is a token wallet, which manages the digital assets of the users and protects the personal data and the access keys. For testing purposes, Metamask can be connected with a local blockchain network by importing the predefined accounts.

\emph{Truffle}~\cite{truffle} is a popular development framework for Ethereum. It provides several tools that facilitate the development process of decentralized applications. Truffle manages smart contracts deployments, automates the smart contracts testing and provides its users with an interactive console, which includes useful commands for smart contracts manipulation.

\emph{Ganache}~\cite{ganache} is a tool provided by Truffle. It creates a local Ethereum blockchain network, which can be used for smart contracts deployment and testing purposes. The users can inspect the issued transactions or events and explore the debugging information. Ganache contains a set of predefined accounts, which can be imported in Metamask using their private keys and the mnemonic code. The information about each account (e.g., address, balance, number of issued transactions) can be also explored.  

\emph{React}~\cite{react} is a JavaScript library used for building the user interface for web applications. We use React and Material UI~\cite{materialui} to create the UI of our decentralized application for rating. 

\emph{Passport}~\cite{passport} is a JavaScript library, which provides a wide range of authentication strategies that can be embedded into a web application. Our solution incorporates an authentication mechanism, which makes use of these strategies.

\section{A decentralized application for rating}
\label{sec:tool}

\noindent
\begin{figure}[t]
    \centering
    \includegraphics[scale=.2]{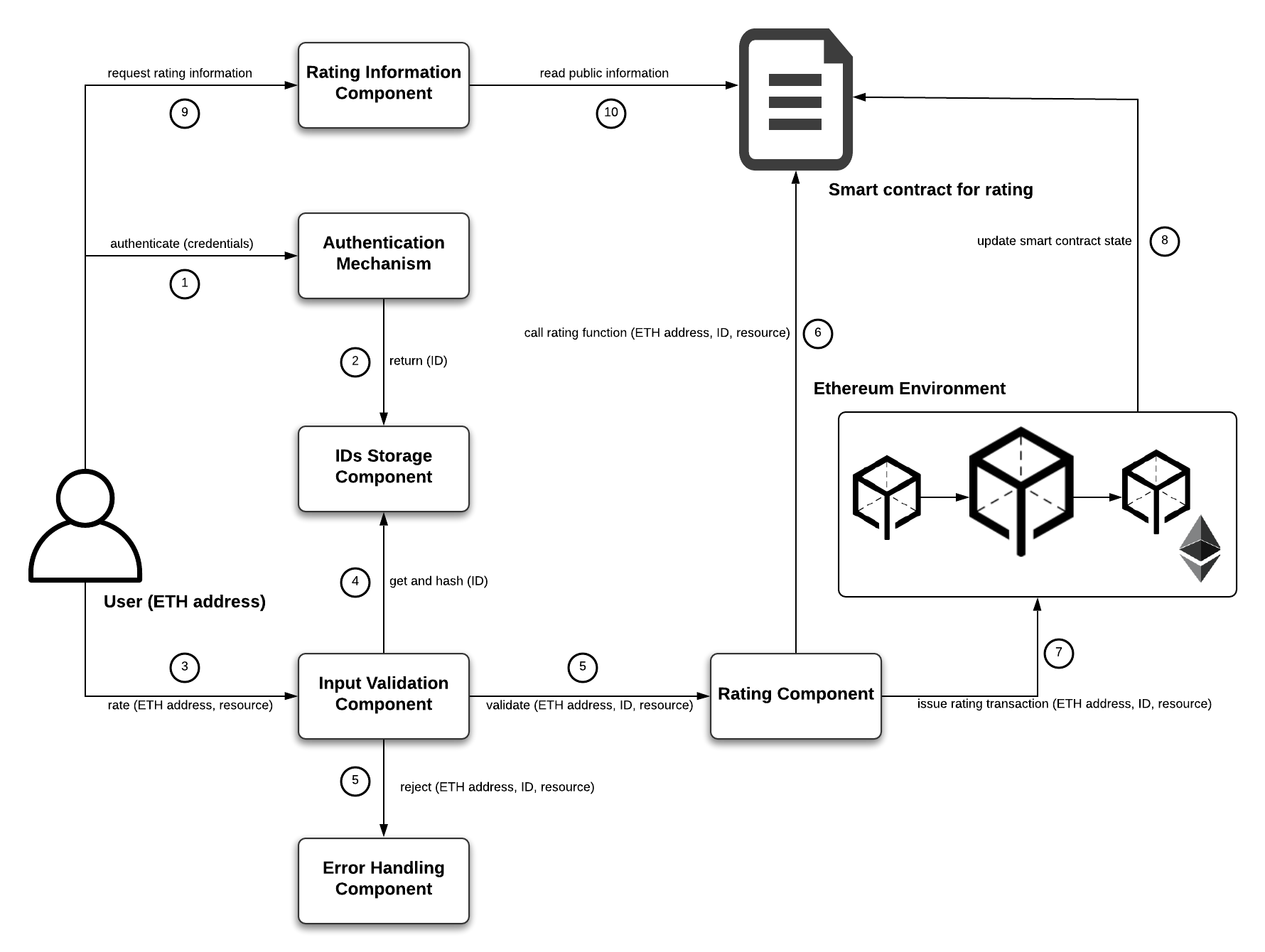}
    \caption{A use case diagram for the decentralized App for Rating. The user interacts with the smart contract using several components: an authentication mechanism, an input validation component, and a component that sends the rating to the smart contract. A query-like component is used to retrieve data from the smart contract. The order of the steps to perform is given by the numbered labels.}
    \label{fig:usecasediagram}
\end{figure}

In this section we present a general workflow of the proposed decentralized application for rating. The application is divided into several components, which have specific roles (e.g., authentication, validation, storage, error handling and rating). The diagram in Figure~\ref{fig:usecasediagram} illustrates the main use cases of the application and highlights the steps of the rating process using numbered labels. We explain these steps in the next paragraphs.

First, the users need to authenticate (step 1 in Figure~\ref{fig:usecasediagram}) with one of the available providers (e.g., Google, GitHub and Spotify). If the user provides valid credentials then the authentication process succeeds and an ID of the user is returned (step 2). The IDs are used when users initiate a rating operation. 

After authentication, the users can rate a diversity of internet resources (step 3) using their IDs (step 4). In order to be recorded, the ratings need to be valid. The application contains a component responsible for input validation. Each time an user initiates a rating operation, the validation component checks if a set of conditions (e.g., if the rated resource is truly hosted by a specific provider or if the user has already rated a specific resource) is met (step 5). If the requirements are satisfied, the rating process can continue, otherwise an error is thrown.

Finally, the rating component is used to interact with the smart contract (step 6). The smart contract exposes a function that handles the rating, that has to be called only with valid arguments. The function creates a new transaction (step 7), sends it across the network and the state of the smart contract is eventually updated (step  8). 

There is also a component responsible for retrieving the rating data (steps 9 and 10) from the smart contract that informs the users with respect to the updated rating history.

\subsection{DApp components}

In the current section, we will focus our attention on the main components of the decentralized application: the smart contract, the authentication mechanism and the user interface. We will briefly discuss the role of each component within the application.
 
\begin{enumerate}
    
\item The \emph{smart contract} encapsulates the logic of the rating process. It exposes a public function that is called each time an user initiates a rating operation. This function handles accurately different use cases and alters the state of the smart contract when a rating operation succeeds. The smart contract includes several data structures that keep track of the rating history. We do not perform any additional checks in the smart contract code, since this could increase the execution cost.
    
\item The \emph{authentication mechanism} guarantees the fact that users have valid accounts. Since the users are uniquely identified, the attempts to rate multiple times a specific resource are easily detected. In this way, the fraudulent tendencies are reduced and the users are provided with a trustworthy rating experience. 

\item The \emph{user interface} provides the users with an intuitive visual experience of the rating process and contains three sections: the authentication section, the rating section and the data visualisation section. The user interface accesses the state of the smart contract and retrieves the rating history in an asynchronous manner. The extracted information can be used not only for the data visualisation section, but also for error handling within the rating section.

\end{enumerate}

\subsection{Smart contract for rating}

We designed a smart contract which simulates the behaviour of a \emph{like/dislike-based} rating system. We developed the smart contract in a Solidity environment within the Remix platform. The smart contract allows the users to rate positively or negatively different kinds of media contents. The behaviour of the smart contract does not change, no matter which one is the content provider. The users and the rated resources are uniquely identified and the ratings consist of tuples of the following form: 

\begin{center}
$\langle$ a user identifier, a resource identifier, a boolean value $\rangle$,
\end{center}

\noindent
where the boolean flag represents a positive or a negative rating. The logic of the smart contract was simplified, since a part of the conditions and requirements are handled directly at the application level. This simplification process has no impact on the accuracy of the rating operations, and it reduces considerably the execution costs, which is one of our goals.

\subsubsection{Data structures.}

The smart contract contains a set of data structures which record the rating operations. In this section, we will provide a brief description of the used data structures. The identifiers for users are of type {\tt bytes32} and the identifiers for resources are of type {\tt string}.

\begin{enumerate}

\item {\tt resourceRating} is a {\tt struct} used to monitor the number of likes and dislikes for a specific resource.
\item {\tt resourcesInformation} represents a {\tt mapping}. The structure records the number of likes and dislikes for each rated resource. 
\item {\tt resources} is an {\tt array} data structure. This component stores all the resources that were rated.
\item {\tt ratedResources} represents a {\tt mapping}. It contains tuples of user IDs and boolean values, where the boolean values indicate if a resource was previously rated or not.
\item {\tt ratingsInformation} is a {\tt mapping} data structure. It records a tuple of the following form: user, resource, rating value. This data structure is helpful, because it provides information with respect to the rating history. If an user attempts to rate multiple times (e.g., positively or negatively) a specific resource, the rating operation will be rejected, since this data structure contains the proof of the previous rating operation.
\item {\tt usersToResources} is a similar to the data structure which was previously described. The purpose of this {\tt mapping} is to record if a resource was previously rated by a specific user. The application can handle several error cases based on the information stored within this data structure.

\end{enumerate}

These data structures are used within the rating function exposed by the smart contract. If a rating operation succeeds, these data structures are updated. The state of the smart contract can be accessed externally, since all these data structures are publicly available.

\subsubsection{Smart contract functions.} In this section, we will focus our attention on the functionalities exposed by the smart contract. We will enumerate a set of possible use cases, which are handled by the smart contract.

\begin{enumerate}
    
    \item The users can rate the resources positively or negatively (e.g., by giving likes or dislikes).
    \item The users can change their rating options for the previously rated resources. In this case, the number of likes and dislikes of the resources updates according to the current rate (e.g., if the initial rate is positive and the current rate is negative, the number of likes decreases by one and the number of dislikes increases by one). 
    \item If an user intends to rate a resource that was not previously included in the rating history, the resource is recorded and its number of likes or dislikes, initially 0, updates to 1.
    \item If an user intends to rate a resource that was previously recorded, the number of likes or dislikes increases by one, with respect to the current rate.
    
\end{enumerate}

The {\tt rate} function (Listing 1.1) handles accurately each possible use case of a \emph{like/dislike-based} rating system. Based on the provided arguments ({\tt \_cred} -- the user identifier, {\tt \_res} -- the resource identifier and {\tt \_vote} -- the rate value), this function updates the state of the smart contract each time a new transaction is issued. The resource is plainly recorded within the smart contract, but the identities of the users are protected, since their IDs are previously hashed.

\begin{center}
  \lstset{%
    caption=The implementation for the {\tt rate} function.,
    basicstyle=\ttfamily\scriptsize\bfseries,
    frame=tb
  }
\begin{lstlisting}
function rate(
    bytes32 _cred,
    string memory _res,
    bool _vote
) public {
    if (usersToResources[_cred][_res] == true) {
        if (
            ratingsInformation[_cred][_res] == true &&
            _vote == false
        ) {
            ratingsInformation[_cred][_res] = false;
            resourcesInformation[_res].likes -= 1;
            resourcesInformation[_res].dislikes += 1;
        }
        if (
            ratingsInformation[_cred][_res] == false &&
            _vote == true
        ) {
            ratingsInformation[_cred][_res] = true;
            resourcesInformation[_res].likes += 1;
            resourcesInformation[_res].dislikes -= 1;
        }
    } else {
        if (ratedResources[_res] == false) {
            ratedResources[_res] = true;
            resources.push(_res);
            usersToResources[_cred][_res] = true;
            if (_vote == true) {
                resourcesInformation[_res] = resourceRating(1, 0);
                ratingsInformation[_cred][_res] = true;
            } else {
                resourcesInformation[_res] = resourceRating(0, 1);
                ratingsInformation[_cred][_res] = false;
            }
        } else {
            usersToResources[_cred][_res] = true;
            if (_vote == true) {
                resourcesInformation[_res].likes += 1;
            } else {
                resourcesInformation[_res].dislikes += 1;
            }
        }
    }
}
\end{lstlisting}
\end{center}

In addition to the {\tt rate} function, the smart contract exposes several helper functions:
\begin{enumerate}
    \item {\tt getResourceInformation} returns the number of likes and dislikes associated with a specific resource. The resource identifier is given as argument to this function. 
    \item {\tt getNumberOfRatedResources} returns the total number of rated resources.
    \item {\tt getRatedResource} returns the identifier of a resource based on its index in the list of rated resources.
\end{enumerate}

They are called at the application level for retrieving the rating history or other useful information.

\subsection{Authentication mechanism}

The decentralized application includes an authentication mechanism. In order to rate media contents, the users need to use the credentials from a specific provider (e.g., YouTube, Spotify, GitHub). Even if the users are required to use their credentials, the ID provided by the authentication mechanism is not plainly stored in the smart contract. The credentials used in the rating process are previously hashed with {\tt md5}. In this way, the one cannot find the credentials by simply inspecting the blockchain.
 This authentication mechanism ensures the fact that users have valid accounts. We are interested in this because we need to make sure that the rating process is trustworthy w.r.t. existing accounts. 
 
Our approach is efficient in preventing multiple rating situations. For instance, a smart contract which records the ratings based on the Ethereum addresses of the users is not resistant to multiple rating attempts. An user can generate multiple Ethereum addresses and rate a specific resource using those addresses. In this way, the ratings can be easily tricked. 

The advantage of our solution is precisely this additional authentication layer. Even so, the users can rate multiple times a specific resource, but they need to have active accounts on the platform that share the resource. This is more difficult to achieve than the generation of Ethereum addresses.
We actually reuse existing algorithms for detecting and removing fake accounts currently implemented by the media platforms.

For the authentication mechanism we used Passport, a JavaScript library which provides a variety of authentication strategies. Currently, in our application we provide support for Google, GitHub and Spotify, but other strategies can be easily added. 

\subsection{User Interface}   

The decentralized application offers an intuitive user experience. Besides the capacity of the smart contract to handle the rating process, some operations and conditions are included directly in the application. The user interface is divide into three main sections: authentication, a section for rating and  data visualisation.

\begin{center}
\begin{figure}[t]
    \centering
    \includegraphics[scale=.38]{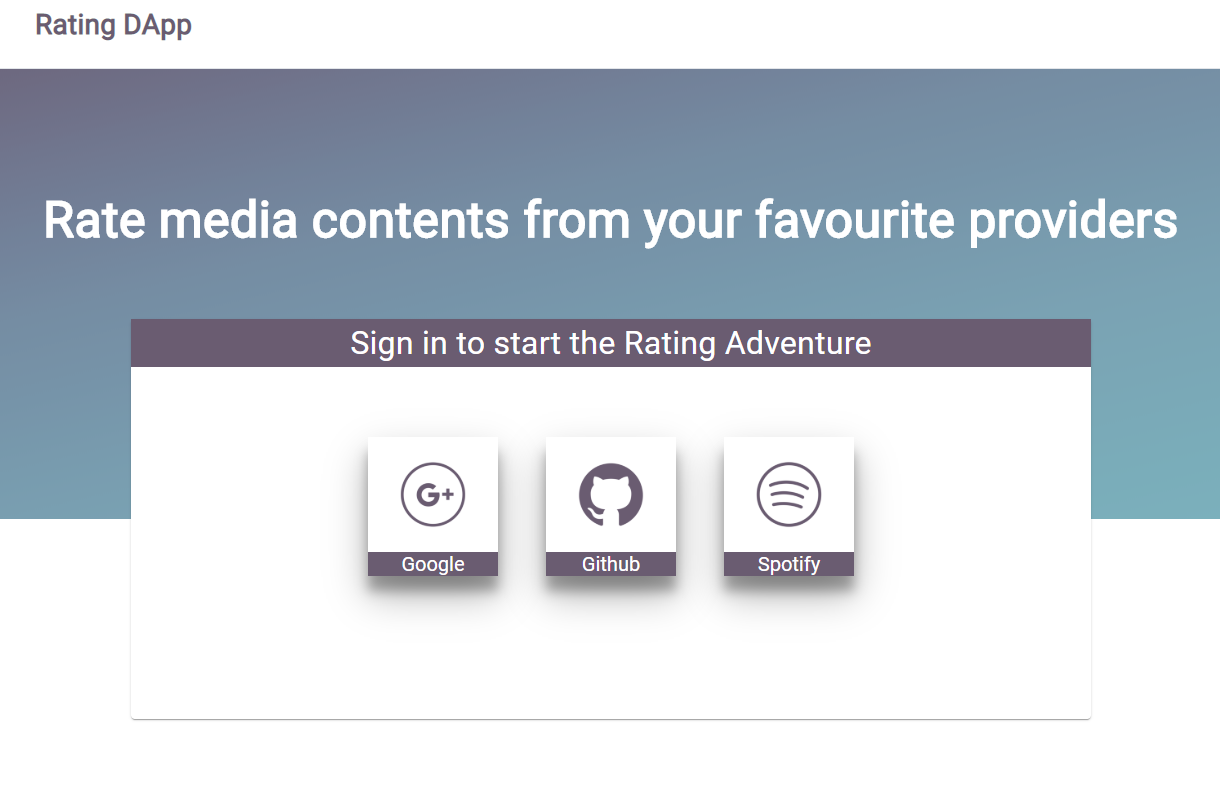}
    \caption{Authentication page where the users can select the platform to sign in. In order to connect with one of the existing log in options, the users should have a valid account on the chosen platform.}
    \label{fig:homepage}
\end{figure}
\end{center}

In the authentication section, the users can choose one of the available login strategies (e.g., Google, GitHub and Spotify). Once the authentication mechanism issues the IDs, the users can start to rate resources. 

The rating section is more complex, because it encapsulates the modules for input validation and error handling. The users need to provide an URL of the resource and to select a rating option (e.g., like or dislike). Before we call the rating function from the smart contract, we perform two input validation steps:

\begin{enumerate}
    \item We check the provenience of the resource;
    \item We check the rating history (i.e., if the user rated the resource with the same rating value, then the user cannot rate it again);
\end{enumerate}

If one of the situations below occurs, the error handling module throws some errors: 
\begin{enumerate}
    \item If the user intends to rate an invalid resource, then the error handling module returns a notification message with the following content: \emph{Invalid resource.}
    \item If the user attempts to rate multiple times the same resource, then the error handling module returns a notification message with the following content: \emph{Multiple ratings for the same resource are not allowed.}
\end{enumerate}
 
If all requirements are met, the rating process continues and a Metamask window will inform the user regarding the rating costs. When the user confirms the transaction, the rating function is called and the state of the smart contract will be modified by the miners in the Ethereum network. The URL-based resources are validated using API calls to the services that share the resources. The state of the smart contract is accessed through reading operations and the rating history can be user to determine if an user rated previously a specific resource.

The data visualisation section accesses the state of the smart contract and the users can inspect directly in the application their rating history. The interactions with the smart contract are asynchronous and it is a matter of seconds until the updated information is displayed within the application. 

We designed the decentralized application using React and Material UI (i.e., JavaScript libraries which provide a variety of predefined UI components).

\begin{figure}[t]
    \centering
    \includegraphics[scale=.39]{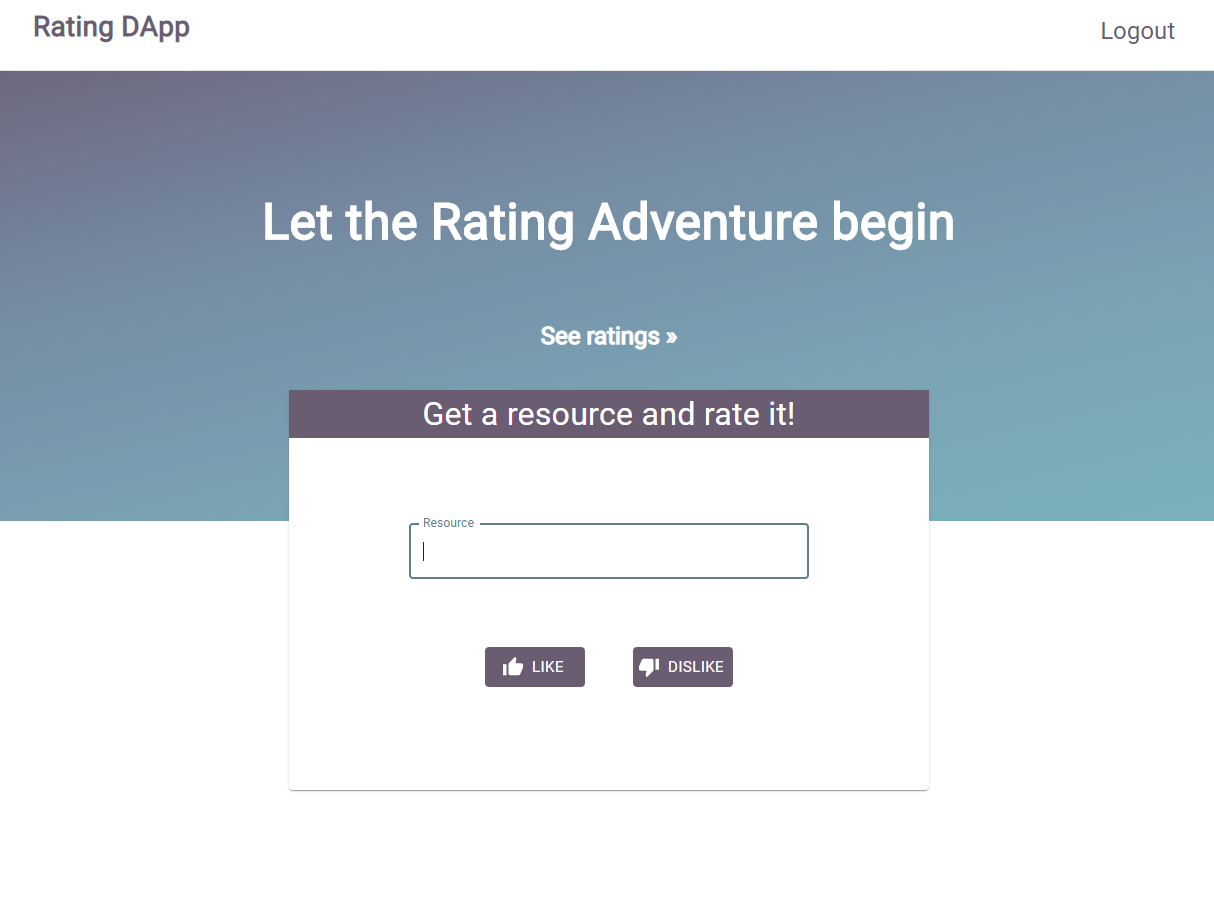}
    \caption{The rating page. The users can complete the form with an URL-based resource and rate it positively or negatively by clicking one of the two specialised buttons. The users can inspect their rating history (e.g., the users have access to a table which contains the rated resources and their associated number of likes/dislikes) by clicking the \emph{See ratings} link. }
    \label{fig:ratingpage}
\end{figure}

\section{Evaluation/Experiments}
\label{sec:tooleval}

In this section we evaluate all the costs required by our decentralized application. There are several categories of costs: the cost to deploy the application, the costs to maintain the smart contract and the costs supported by users for rating the resources. We will present our approaches and their impact on the execution costs.

Our initial attempt was to encapsulate the entire logic within the smart contract. Since our rating system aims to avoid the multiple voting problem and to validate the resource provenience, the smart contract gained in complexity. This led to increased execution costs and made our initial attempt unfeasible. 
Moreover, smart contracts can not access information outside the network. Since we needed confirmations that the rated resources are shared by specific providers, we initially used an oracle (i.e, a service that verifies real world data and submits the collected information to the smart contract). We used Provable~\cite{provable} and Chainlink~\cite{chainlink} to validate the arguments for the rating function. This approach led to different kinds of problems. First of all, the deployment costs and the costs supported by users for the rating operations increased considerably, because the off-chain requests require extra fees. The off-chain request involved also the usage of custom cryptocurrencies (e.g., LINK for Chainlink oracle). 

In order to stimulate the interest for the proposed rating system, we tried to minimize the costs\footnote{The costs involved by the smart contract deployment and the costs for a rating operation with and without off-chain interaction were observed in a testing environment provided by Ganache.}. 
Based on the observation that off-chain interactions are far more suitable, we moved the authentication, input validation and other side logic outside the smart contract.
The complexity of the smart contract is now considerably reduced. Except the minimal information (IDs, resources and their associated ratings) required to keep the ratings in a decentralized manner, the rest of the logic is directly handled at the application level. Thus, the off-chain dependencies were removed from the smart contract.

In Table~\ref{tbl:costs}, we present a brief comparison between the versions of the smart contract that we have developed. The ones that use external providers (oracles) like Provable or Chainlink have an unacceptable execution cost for rating. The simple smart contract with no external providers has significantly lower execution costs per rating operation. This is the one we have currently implemented\footnote{The code is available on GitHub:\url{https://github.com/buterchiandreea/rating-dapp/blob/master/contracts/Rating.sol}}.
\begin{center}
\begin{table}[htp]
    \centering
    \begin{tabular}{ |p{3.5cm}||p{3cm}|p{3cm}|| }
    \hline
    \multicolumn{3}{ |c| }{\bf Costs analysis } \\
    \hline
    {\bf Smart Contract version} & {\bf \hfil Deployment} & {\bf \hfil Rating operation} \\
    \hline
    Simple smart contract (no external providers) & \hfil10 \$ & \hfil 0.2 \$ \\ 
    \hline
    Provable & \hfil10 \$ & \hfil 2 \$ \\
    \hline
    Chainlink & \hfil10 \$ & \hfil 2-8 \$ \\
    \hline
    \end{tabular}\vspace*{2ex}
    \caption{A comparison between various smart contract versions that we experimented. The table shows the costs for deployment and rating operations. Note that the versions of the smart contract that use external providers (oracles) like Provable or Chainlink have an unacceptable execution cost for rating.}
    \label{tbl:costs}
\end{table}
\end{center}

The codebase is available at \url{https://github.com/buterchiandreea/rating-dapp}. The Github repository contains the source code and installation instructions. The smart contract code is written in {\tt Rating.sol} under the {\tt contracts} folder, while the UI is mainly in the {\tt client} folder. 

\section{Conclusions}
\label{sec:conclusions}

\begin{figure}[t]
    \centering
    \includegraphics[scale=.19]{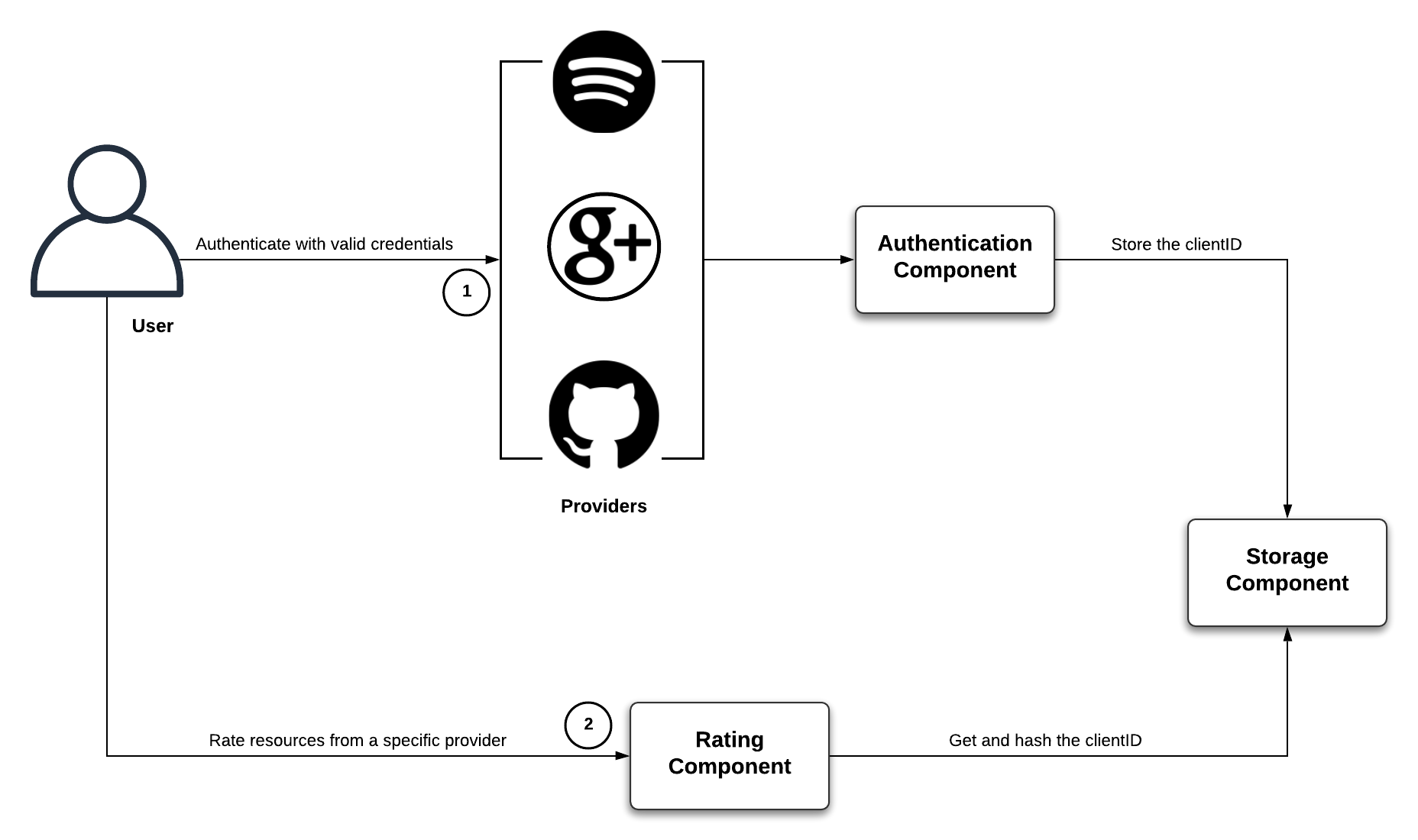}
    \caption{Authentication and rating. The users can not rate a resource until the authentication process is completed. The ratings are recorded based on the {\tt clientID}, which is returned after the authentication succeeds. In this way, we constraint the number of votes to maximum 1 for each account.}
    \label{fig:authentication}
\end{figure}

The solution we propose in this paper is a decentralized application that stores the rating in the blockchain using a simple smart contract. In order to rate an internet resource, users need accounts on the platform that shares that resource. Obviously, their accounts or login information are not stored in the blockchain, but their hashed value is used to limit the number of votes to a maximum of 1 per account. Note that this approach keeps the  benefits brought by the algorithms for eliminating or detecting fake accounts currently implemented in the existing media platforms. In this context, our rating application cannot accept more fake ratings than the existing approaches, as shown in Figure~\ref{fig:authentication}. The transparency and decentralisation are guaranteed by the use of the blockchain. 

To summarise, our approach provides a decentralized solution to keep ratings for internet resources identified by an URL. The approach is based on the blockchain technology that provides anonymity and immutability of data besides decentralisation. We use authentication with third parties in order to provide an inferior limit to the number of votes. The limit is the number of accounts controlled by an user. 

\paragraph{Future work.} There are at least two new features than we intent to add to our decentralized application. First, we want to add more extensions, so that users can authenticate with other accounts. Currently we support only Google, Spotify, and GitHub, but our architecture allows any extension that is compatible with Passport.
Another improvement is related to the various types of ratings. At this point, our application only allows users to rate resources using a two choice (positive/negative) option. However, other rating systems allow scores or systems based on stars. Adding reviews is also an interesting feature, but this requires a very strong analysis on the costs involved to keep the reviews on the blockchain. An interesting idea would be not to keep the reviews on the blockchain, but only a hashed value of the review so that it can be easily checked if a review retrieved from a different resource is indeed the one whose hash is stored in the blockchain.

\newpage

\bibliography{refs}

\begin{thebibliography}{10}

\bibitem{Ayed2017ACS}
Ahmed~Ben Ayed.
\newblock A conceptual secure blockchain based electronic voting system.
\newblock {\em International Journal of Network Security \& Its Applications},
  9:01--09, 2017.

\bibitem{10.1007/978-3-642-02627-0_5}
Thomas Bocek, Dalibor Peric, Fabio Hecht, David Hausheer, and Burkhard Stiller.
\newblock Peervote: A decentralized voting mechanism for p2p collaboration
  systems.
\newblock In Ramin Sadre and Aiko Pras, editors, {\em Scalability of Networks
  and Services}, pages 56--69, Berlin, Heidelberg, 2009. Springer Berlin
  Heidelberg.

\bibitem{eth}
Vitalik Buterin.
\newblock Ethereum : A next-generation smart contract and decentralized
  application platform, 2013.

\bibitem{chainlink}
Chainlink.
\newblock Chainlink docs.
\newblock https://chain.link/, 2020.

\bibitem{solidity}
Ethereum Foundation.
\newblock Solidity docs.
\newblock https://solidity.readthedocs.io/en/v0.6.10/, 2020.

\bibitem{vyper}
Ethereum Foundation.
\newblock Vyper docs.
\newblock https://vyper.readthedocs.io/en/stable/, 2020.

\bibitem{hyperledger}
Linux Foundation.
\newblock Hyperledger docs.
\newblock https://www.hyperledger.org/, 2020.

\bibitem{10.1007/978-981-10-7605-3_50}
Jen-Ho Hsiao, Raylin Tso, Chien-Ming Chen, and Mu-En Wu.
\newblock Decentralized e-voting systems based on the blockchain technology.
\newblock In James~J. Park, Vincenzo Loia, Gangman Yi, and Yunsick Sung,
  editors, {\em Advances in Computer Science and Ubiquitous Computing}, pages
  305--309, Singapore, 2018. Springer Singapore.

\bibitem{mci/Khader2012}
Dalia Khader, Ben Smyth, Peter Y.~A. Ryan, and Feng Hao.
\newblock A fair and robust voting system by broadcast.
\newblock In Manuel~J. Kripp, Melanie Volkamer, and Rüdiger Grimm, editors,
  {\em 5th International Conference on Electronic Voting 2012 (EVOTE2012)},
  pages 285--299, Bonn, 2012. Gesellschaft für Informatik e.V.

\bibitem{8603050}
D.~{Khoury}, E.~F. {Kfoury}, A.~{Kassem}, and H.~{Harb}.
\newblock Decentralized voting platform based on ethereum blockchain.
\newblock In {\em 2018 IEEE International Multidisciplinary Conference on
  Engineering Technology (IMCET)}, pages 1--6, 2018.

\bibitem{kovan}
Kovan.
\newblock Kovan docs.
\newblock https://kovan-testnet.github.io/website/, 2020.

\bibitem{metamask}
Metamask.
\newblock Metamask docs.
\newblock https://docs.metamask.io/guide/, 2020.

\bibitem{bitcoin}
Satoshi Nakamoto.
\newblock Bitcoin: A peer-to-peer electronic cash system.
\newblock {\em Cryptography Mailing list at https://metzdowd.com}, 03 2009.

\bibitem{passport}
Passport.
\newblock Passport docs.
\newblock http://www.passportjs.org/docs/, 2020.

\bibitem{provable}
Provable.
\newblock Provable docs.
\newblock https://docs.provable.xyz/, 2020.

\bibitem{react}
React.
\newblock Github react.
\newblock https://github.com/facebook/react/, 2020.

\bibitem{remix}
Remix.
\newblock Remix docs.
\newblock https://remix-ide.readthedocs.io/en/latest/, 2020.

\bibitem{rinkeby}
Rinkeby.
\newblock Rinkeby docs.
\newblock https://www.rinkeby.io/, 2020.

\bibitem{ropsten}
Ropsten.
\newblock Github ropsten.
\newblock https://github.com/ethereum/ropsten, 2020.

\bibitem{ganache}
Truffle Suite.
\newblock Ganache docs.
\newblock https://www.trufflesuite.com/docs/ganache/overview, 2020.

\bibitem{truffle}
Truffle Suite.
\newblock Truffle docs.
\newblock https://www.trufflesuite.com/docs, 2020.

\bibitem{materialui}
Material UI.
\newblock Material ui docs.
\newblock https://material-ui.com/, 2020.

\bibitem{wood2014ethereum}
Gavin Wood.
\newblock Ethereum: a secure decentralised generalised transaction ledger.
\newblock https://gavwood.com/paper.pdf, 2014.

\end{thebibliography}
\bibliographystyle{plain}

\end{document}